\documentstyle[12pt]{article}
\renewcommand{\theequation}{\thesection.\arabic{equation}}
\renewcommand{\thesubsection}{\thesection.\arabic{subsection}}

\newlength{\extraspace}
\newlength{\extraspaces}
\newcounter{dummy}

\newcommand{\be}{\begin{equation}
\addtolength{\abovedisplayskip}{\extraspaces}
\addtolength{\belowdisplayskip}{\extraspaces}
\addtolength{\abovedisplayshortskip}{\extraspace}
\addtolength{\belowdisplayshortskip}{\extraspace}}
\newcommand{\ee}{\end{equation}}

\newcommand{\ba}{\begin{eqnarray}
\addtolength{\abovedisplayskip}{\extraspaces}
\addtolength{\belowdisplayskip}{\extraspaces}
\addtolength{\abovedisplayshortskip}{\extraspace}
\addtolength{\belowdisplayshortskip}{\extraspace}}
\newcommand{\ea}{\end{eqnarray}}

\newcommand{\baa}{
\addtocounter{equation}{1}
\setcounter{dummy}{\value{equation}}
\setcounter{equation}{0}
\renewcommand{\theequation}{\thesection.\arabic{dummy}\alph{equation}}
\begin{eqnarray}
\addtolength{\abovedisplayskip}{\extraspaces}
\addtolength{\belowdisplayskip}{\extraspaces}
\addtolength{\abovedisplayshortskip}{\extraspace}
\addtolength{\belowdisplayshortskip}{\extraspace}}
\newcommand{\eaa}{
\end{eqnarray}
\setcounter{equation}{\value{dummy}}
\renewcommand{\theequation}{\thesection.\arabic{equation}}}

\newcommand{\ban}{\begin{eqnarray*}
\addtolength{\abovedisplayskip}{\extraspaces}
\addtolength{\belowdisplayskip}{\extraspaces}
\addtolength{\abovedisplayshortskip}{\extraspace}
\addtolength{\belowdisplayshortskip}{\extraspace}}
\newcommand{\ean}{\end{eqnarray*}}

\newcommand{\newsection}[1]{
\vspace{16mm}
\pagebreak[3]
\addtocounter{section}{1}
\setcounter{equation}{0}
\setcounter{subsection}{0}
\setcounter{footnote}{0}
\begin{flushleft}
{\large\bf \thesection. #1}
\end{flushleft}
\nopagebreak
\smallskip
\nopagebreak}

\newcommand{\newsubsection}[1]{
\vspace{10mm}
\pagebreak[3]

\addtocounter{subsection}{1}
\addcontentsline{toc}{subsection}{\protect
\numberline{\arabic{section}.\arabic{subsection}}{#1}}
\noindent{\sc \thesubsection. #1}
\nopagebreak
\vspace{2mm}
\nopagebreak}

\newcommand{\ie}{{\it i.e.}}

\newcommand{\tr}{{\rm tr}}
\newcommand{\is}{\! & \! = \! & \!}

\newcommand{\nonu}{\nonumber \\[1.5mm]}

\newcommand{\ra}{\rightarrow}

\newcommand{\half}{{\textstyle{1\over 2}}}

\newcommand{\X}{{\mbox{\footnotesize $X$}}}
\newcommand{\PP}{{\mbox{\footnotesize $P$}}}

\newcommand{\vac}{|0\rangle}

\newcommand{\del}{\partial}

\newcommand{\ka}{\kappa}
\newcommand{\xx}{{\tt x}}
\newcommand{\rhohat}{\hat{\rho}}

\newcommand{\Tau}{u}
\newcommand{\Time}{v}

\begin{document}
\addtolength{\baselineskip}{.3mm}
\input epsf

\thispagestyle{empty}

\vspace{-1cm}

\begin{flushright}
{\sc CERN-TH}.7142/94\\
{\sc PUPT}-1441\\
January 1994
\end{flushright}
%\vspace{.1cm}

\begin{center}
{\large\sc{Black Hole Evaporation %\\[3mm]
and Quantum Gravity.}}\footnote{Based on talks given by E. V.
at the Trieste Spring School on Strings, Gravity and Gauge Theory, Trieste,
April 1993,
by E. V. and K. S. at Strings '93,
Berkeley, May 1993 and by H. V. at the Conference on Quantum Aspects of
Black Holes, Santa Barbara, June 1993.}\\[10mm]
{\sc Kareljan Schoutens, Herman Verlinde}\\[3mm]
{\it Joseph Henry Laboratories\\[2mm]
Princeton University, Princeton, NJ 08544} \\[.3cm]
{ and}\\[.3cm]
{\sc Erik Verlinde}\\[3mm]
{\it TH-Division, CERN\\[2mm]
CH-1211 Geneva 23\\[2mm]
and\\[2mm]
Institute for Theoretical Physics\\[2mm]
University of Utrecht\\[2mm]
P.O. BOX 80.006, 3508 TA Utrecht}\\[13mm]

{\sc Abstract}
\end{center}

\vspace{-1mm}
\noindent
In this note we consider some consequences of quantum gravity
on the process of black hole evaporation. In particular, we will
explain the suggestion by 't Hooft that quantum gravitational interactions
effectively exclude simultaneous measurements of the Hawking radiation and
of the matter falling into the black hole. The complementarity of these
measurements is supported by the fact that the commutators between the
corresponding observables can be shown to grow uncontrollably large.
The only assumption that is needed to obtain this result is that the
creation and annihilation modes of the in-falling and out-going matter act
in the same Hilbert space. We further illustrate this phenomenon in the
context of two-dimensional dilaton gravity.

\newpage

\newsection{Introduction.}

The black hole evaporation phenomenon can be viewed as a consequence of the
fact that the horizon of a black hole on the one hand forms a surface of
infinite red-shift, while on the other hand it represents a perfectly
regular part of space-time. Any out-going wave that reaches an asymptotic
observer with a finite frequency corresponds %, when propagated back in time,
to an exponentially high frequency mode near the horizon, and the reasonable
assumption that these ultra-high energy modes are in their ground state was
used by Hawking to show that the asymptotic observer will see thermal
radiation \cite{hawking}.
Soon after this discovery, Hawking made the remarkable suggestion that
the resulting black hole evaporation process will inevitably
lead to a fundamental loss of quantum coherence. The mechanism by which
the quantum radiation is emitted indeed appears to be insensitive to
the detailed history of the black hole, and thus it seems
hard to imagine how one can prevent that information gets forever lost
to an outside observer.

The only possibility for maintaining quantum coherence, it seems,
would be if quantum gravity somehow leads to non-local effects that
gradually bring out this information in the form of subtle
correlations in the out-going radiation. Superficially, however,
the evaporation process for large black holes involves only
physics at low energies. After all, the process entirely takes place
within space-time regions where both the curvature and the energy-momentum
flux of the radiation remain small everywhere. It is therefore often
argued that any strong quantum gravitational effects take place either
too late or too far behind the horizon to provide a possible
mechanism for a complete information transport to the outside world.

This last reasoning however is incomplete. Namely, as has been emphasized by
several authors (see e.g. \cite{thooft} and \cite{jacobson}), it does not
take into account the important fact that the exponential red-shift effect
associated with the black hole horizon
leads to a breakdown of the usual separation of length scales.
In a certain way, this red-shift effectively works as a magnifying glass
that makes
the consequences of the short distance, or rather, high energy physics near
the horizon visible at larger scales to an asymptotic observer.
Direct examination of Hawking's original derivation (or any later one) of
the black hole emission spectrum indeed shows that one inevitably needs
to make reference to particle waves that have arbitrarily high frequency
near the horizon  as measured in the reference frame of the in-falling matter.
While this point has been noted by many authors, it
is usually put aside with the argument that, since there are no physical
ultra-high energy particles running along the horizon, one does not need
to know anything about their physics except how to describe their local
vacuum state.
%Furthermore, even though the frequencies may get very large
%in the reference frame of the in-falling observer, it seems one can always
%go to a boosted frame near the horizon in which the frequency of the waves
%remain small.
However, after one realizes that the frequencies involved here
are so large that the corresponding energies exceed any macroscopic
mass by an exponentially growing factor, it becomes in fact far from
obvious that quantum gravitational effects remain small.
At these frequencies it may indeed no longer be appropriate to think
of particles as being superimposed on top of some classical
background geometry.

In this respect it is important to note that, while
often we think of quantum gravity as relevant only to the physics at
sub-planckian distances, at these ultra-high frequencies quantum
gravitational effects can in principle take macroscopic proportions.
Only if as an asymptotic observer we would restrict our observations
to extremely low frequencies, such that they remain
reasonably small when propagated back to the horizon, we can to a very
good approximation work with the classical geometry. If instead we
measure out-going particles of moderate frequencies at infinity,
then the history of these modes must involve geometries that can be very
different from the classical one.

As has been emphasized by 't Hooft, this fact may lead to large
deviations from our semi-classical intuition. In particular, he has
suggested that, when one wants to simultaneously consider the
observations of an asymptotic observer and those of an in-falling observer,
such measurements will in general involve observables
whose commutators will grow uncontrollably large. Hence these simultaneous
measurements are essentially forbidden: they are {\it complementary}
in the usual sense of quantum mechanics. If this suggestion is
true, then it will obviously have very important consequences.

An apparent
weakness in the argumentation of 't Hooft, however, is that this
complementarity between the in-falling and asymptotic observers appears to
follow almost directly from the starting assumption that the black
hole evaporation process should be describable in terms of an $S$-matrix.
This assumption by itself immediately turns the black hole horizon into
an ultra-strong coupling regime, because any $S$-matrix element involving
a generic $out$-state will be very singular in that region.
This very fact, on the other hand, is often used as an argument
to show why the $S$-matrix assumption must be wrong, as it appears to
contradict the fact that horizon should be regular to an
in-falling observer.
It is therefore important to establish whether the same non-trivial quantum
gravitational effects can be derived in a more general framework that is
not based on this $S$-matrix assumption.

As a first small step in this direction, two of us \cite{unpub} and
independently Susskind and collaborators \cite{sussetal} presented
several arguments suggesting that the idea of complementarity can
not be disproven without making reference to Planckian physics.
In this note, however, we would like to go further and indicate how
one can actually {\it derive} the existence of these large commutators
in the standard set-up chosen by Hawking, essentially without making
any further unnecessary assumptions.
The key new ingredient in our discussion will be that
we will take into account the quantum mechanical nature of the matter
forming the black hole.  We will make this step by simply replacing
in Hawking's original formulas the classical in-falling stress-energy by
the corresponding quantum mechanical operator.
One of the surprising consequences of this procedure is that it
automatically includes an important part of the gravitational back
reaction and it also gives some new insight into the question of
energy conservation.

Another very important new effect is that the asymptotic coordinate
system, since it is dynamically determined in terms of the in-falling
matter, becomes {\it operator valued}. In a Heisenberg
picture, this will imply that the observables that measure the asymptotic
radiation will {\it not commute} with the observables associated with the
in-falling matter.
Normally, if we draw a Cauchy surface in a
space-time diagram, like that in figure 1, we expect that all
operators on this surface that are spacelike separated commute with
each other. This assumption is in fact essential in the standard
argumentation that the asymptotic Hilbert space of $out$-modes
is incomplete. However, as we will explicitly show in this note, the
commutators between $out$-modes $\phi_{out}$ and in-going
modes $\phi_{in}$
in figure 1 are non-zero, and in fact will grow extremely
large. It is clear that this result
will require a drastic revision of the standard
semi-classical picture of the evaporation process.

\begin{center}
\leavevmode
\epsfysize=5.5cm
\epsfbox{figu1.eps}
\end{center}
\noindent
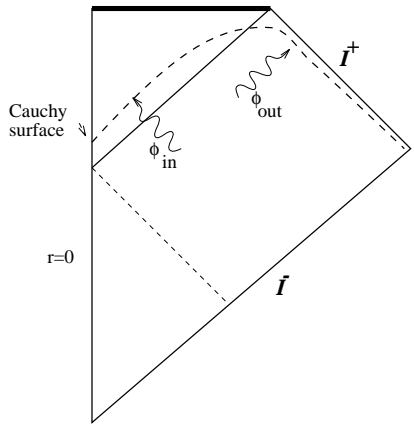
{\small Fig 1. We will show in this note that,
due to the fact that the asymptotic coordinate system
is dynamically determined in terms of the in-falling matter, the
commutator between the out-going modes $\phi_{out}$ and in-falling
modes $\phi_{in}$ will grow uncontrollably large.

\medskip}

In the first part of this note we will present this calculation for
the $s$-wave sector of the $3+1$-dimensional  problem. After reviewing
Hawking's
original derivation we will argue on the basis of energy conservation
that this derivation, as it stands, must be corrected already after
a very short time. We will then summarize the derivation of the
commutator between the $in$ and $out$ fields, and briefly discuss some
implications of this result.
In the second half of these notes we will illustrate this
phenomenon and some possible physical consequences more explicitly in the
context of two-dimensional dilaton gravity. This second part will also
contain as a new result an exact quantum construction of the physical
Hilbert space for an arbitrary number of massless matter
fields. In this model the dynamical nature of the asymptotic coordinate
system follows naturally from the %standard
gravitational dressing of the matter fields.

\newsection{Black Hole Evaporation in the $s$-Wave Sector}

In this section we begin with a short summary of Hawking's original
derivation of the thermal spectrum of the out-going radiation.
To simplify the formulas we restrict our attention to the $s$-wave sector of
a massless field $\phi$.  We introduce two null-coordinates $u$ and
$v$  such that at a large distance $r\ra\infty$  we have $u\ra
t+r$ and $v\ra t-r$.

\newsubsection{Hawking's derivation.}

In his pioneering paper \cite{hawking}, Hawking did not not refer to
the local vacuum state near the horizon, but instead he tried to
establish a direct relation between
the out-going state at future null infinity ${\cal I}^+$ directly
to the in-state at past null infinity ${\cal I}^-$.
Specifically,
he imagined sending a small test-particle backwards in time from
future null infinity ${\cal I}^+$ and letting it propagate all the
way through to ${\cal I}^-$ (see fig 2).
To relate the form of this signal in the two asymptotic regions,  he
then used the free wave equation on the {\it fixed} background geometry
of the collapsing black hole,
while ignoring the effects due to gravitational back reaction.

\begin{center}
\leavevmode
\epsfysize=5.5cm
\epsfbox{figu2.eps}
\end{center}
\noindent
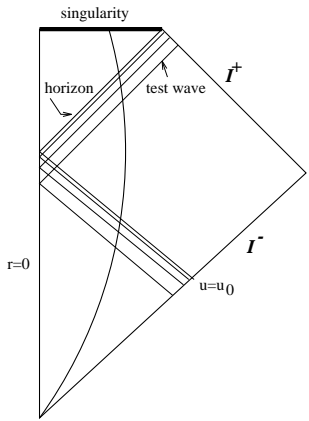
{\small Fig 2. Following Hawking, we construct the out-going state at
${\cal I}^+$ by sending back a test wave and letting it propagate all
the way to ${\cal I}^-$. Note that the frequency of this test wave diverges
near the event horizon.
}

{}From the condition that the field is regular at the origin $r=0$
one deduces that the outgoing $s$-wave $\phi_{out}$ of
the massless scalar field $\phi$ and the corresponding
in-coming wave $\phi_{in}$ are related by a reparametrization
\be\label{repar}
\phi_{in}(\Tau)=\phi_{out}(\Time(\Tau)).
\ee
%that relates the null-coordinate $\Time$ along ${\cal I}^+$ to the
%null-coordinate $\Tau$ along ${\cal I}^-$.
The diffeomorphism $\Time(\Tau)$ typically takes the form
\be
\label{ttau}
\Time(\Tau)=\Tau-4M\log[(\Tau_0-\Tau)/4M],
\ee
where $M$ denotes the black hole mass and $\Tau_0$ the critical in-going
time, \ie\ the location of the in-going light-ray that later will coincide
with the black hole horizon.
Thus an outgoing $s$-wave with a given frequency $\omega$ translates
to an in-signal
\be
\label{asym}
e^{i\omega \Time(\Tau)}=e^{i\omega\Tau} ({\Tau_0-\Tau\over 4M})^{-i4M\omega}
\theta(\Tau_0-\Tau).
\ee
that decomposes as a linear superposition of incoming waves
with very different frequencies.
The out-going modes $b_\omega$ (\ie\ the Fourier coefficients of
$\phi_{out}$) are therefore related to the in-coming modes $a_\xi$
via a non-trivial Bogoljubov transformation of the form
\ba
\label{Bogol}
b_\omega\is\sum_{\xi} \alpha_{\omega\xi} a_\xi + \sum_{\xi}\beta_{\omega\xi}
a^\dagger_\xi\nonu
b^\dagger_\omega \is \sum_{\xi} \beta^*_{\omega\xi} a_\xi
+ \sum_{\xi}\alpha^*_{\omega\xi}
a^\dagger_\xi.
\ea
The coefficients $\alpha_{\omega\xi}$ and $\beta_{\omega\xi}$ of this
transformation are the Fourier transform of the function that appears on
the right-hand side of (\ref{asym}). Up to some irrelevant phase the
Bogoljubov coefficients take the form
\ba
\label{coeff}
\alpha_{\omega\xi} \is \, \sqrt{\omega\over\xi}\,e^{i(\xi-\omega)\Tau_0} e^{\pi
M \omega}\Gamma(1-i4M\omega)
\nonu
\beta_{\omega\xi} \is \, \sqrt{\omega\over\xi}\,e^{i(\xi+\omega)\Tau_0} e^{-\pi
M \omega}\Gamma(1+i4M\omega)
\ea
At this level of approximation, these coefficients are $c$-numbers.

Once this relation between the
$out$-modes and $in$-modes is established, one can for any given
$in$-state find the corresponding $out$-state.
In general, this $out$-state can be represented by a density matrix
$\rho_{out}$, such that
\be
\label{map}
\tr( \rho_{out} {\cal O}(b^\dagger,b) )= \langle in |
{\cal O}(\beta^* a \!+\!\alpha^*a^\dagger,\alpha\, a\!+\!
\beta \, a^\dagger)|in\rangle.
\ee
for all operators $\cal O$ that are constructed from the $b$-modes.
When this map is applied to the in-vacuum $|0\rangle$ one obtains
the Hawking state $\rho_H$. The explicit form of this Hawking state
can be computed explicitly in terms of the coefficients
(\ref{coeff}), and describes a constant flux of thermal radiation
at the Hawking temperature $T_H= {1\over 8\pi M}$.
Notice that the Hawking state also depends on the time $\Tau_0$
at which the black hole horizon forms, but that this dependence only shows up
in the expectation values of operators $\cal O$ carrying non-zero energy.

Since the
transformation
(\ref{Bogol}) is not invertible,
the map $|in\rangle\ra \rho_{out}$ defined in (\ref{map})  maps
{\it pure} $in$-states into {\it mixed} $out$-states.
Indeed, in-coming waves that have support for
$\Tau>\Tau_0$ cannot be re-constructed out of the outgoing
$s$-waves, and thus the out-going creation operators $b^\dagger_\omega$
generate only a subspace ${\cal H}_{out}$ of the complete Hilbert
space ${\cal H}$ generated by the $a^\dagger_\xi$-modes. As long as the
black hole is still there this is not so surprising, because the
total $out$ Hilbert space also contains a sector describing the
matter that has fallen into the black hole.
The important question is however whether the outgoing radiation, after
the black hole has completely evaporated,  is still described by a mixed state
or whether  eventually quantum coherence is restored.
Clearly, this last possibility can only occur if the gravitational
back reaction leads to a drastic modification of the
above semi-classical picture.

\newsubsection{Ultra-high frequencies.}

In the above derivation of the Hawking state we have ignored
the gravitational back reaction of  the test wave on the geometry.
This is commonly believed to be a good approximation, at least for
for macroscopic black holes,  since one only
considers space-time regions in which both the
curvature as well as the expectation value of the stress-tensor are small
compared to the Planck scale. This argument, however,
is in our opinion at best sufficient to show
that the Hawking state correctly describes the global features
of the out-going radiation, such as the average energy-flux. To study the
actual information content of the radiation, on the other hand, one has
to know the {\it entire} quantum state and in this case the behaviour of
one {\it single}
expectation value provides an insufficient criterion for the reliability
of the approximation.

Because the out-going radiation has an (approximately) thermal spectrum with
temperature $T_H= {1\over 8\pi M}$, the typical frequency $\omega$ of an
out-going mode $b_\omega$ is of the order $$\omega\sim {1\over M }$$
in Planck units.
So, as long as $M$ is much larger than the Planck mass,
each Hawking particle carries only a very small
fraction of the total mass of the black hole.
On the other hand,
from (\ref{asym}) we see that the corresponding in-coming modes
are very rapidly oscillating near $\Tau\ra \Tau_0$. This implies that
the calculation of the Hawking state at late time $\Time_1$
 requires one to consider very high in-coming frequencies, typically
of the order of
\be
\label{freq}
\xi\sim  {e^{(\Time_1-\Tau_0)/4M} \over M }.
\ee
Hence, already
after a very short time of the order of $M\log({M})$
after the formation of the black hole, we need to consider
frequencies $\xi$ that are much larger than the mass $M$ of the black hole
itself!
Note that even
for a macroscopic solar-mass black hole this is already after a fraction of
a second!

It is now immediately clear that, due to the appearance of these large
frequencies, the result (\ref{Bogol})-(\ref{coeff})
for the relation between the
$in$ and $out$-modes can not be taken literally. Namely, suppose that we
would trust the transformation (\ref{Bogol}) as an accurate
approximation of the exact operator identification.
Then, physically, it describes how a particle, while
propagating in the (fixed) background geometry
of a collapsing star, changes its energy
from its initial value $\xi$ to a final value $\omega$. In fact,
the quantum mechanical amplitude for this process is given by one
of the Boboljubov coefficients
\be
\label{ampl}
\langle 0| b_\omega a^\dagger_\xi|0\rangle = \alpha_{\omega\xi}\ .
\ee
In general the initial energy $\xi$ is much larger
than the final energy $\omega$, because a particle will lose
almost all its energy while trying to escape from near the horizon
out to infinity.  However, suppose that we do include the back reaction,
then the total energy of the particle-black hole system
should be conserved of course.
We must therefore conclude that all the energy that is
lost by the particle
has gone into the black hole.  So, if in the final state
the mass of the black hole is equal to $M$, the initial
black hole mass must have been $$M_{initial}=M+\omega-\xi,$$
just like in any other physical scattering process.
Now it is clear that for real physical
particles the gravitational back reaction will surely
lead to significant corrections in the amplitude (\ref{ampl}) as soon
as the difference $\xi-\omega$ is of the order of $M$. In this regime the
linear transformation
(\ref{Bogol}) between the $in$ and $out$-modes
no longer gives an accurate description of
particle-black hole scattering, and must therefore be corrected.
In our opinion this also means that to correctly describe the
Hawking process %after a time $2M\log(M)$ after the black hole formation
one should use a modified version of the Bogoljubov transformation that
includes gravitational corrections.

While it should be evident from this that the formula (\ref{Bogol}),
when interpreted as an operator identification, needs to be corrected,
it is perhaps not so clear that these corrections will also lead to real
modifications in the calculation of the Hawking state. One could indeed
argue that in this calculation
one doesn't quite need the relation (\ref{Bogol}) to its full strength,
since one is only interested in propagating an $in$-state that looks
like the {\it vacuum} near $\Tau = \Tau_0$. In particular, it would appear
therefore that one can use the (approximate) Lorentz invariance of this local
vacuum state to drastically reduce the problem of the exploding
frequencies. However, this mechanism is clearly insufficient to
completely eliminate the problem, because this would essentially
require that Lorentz invariance is an {\it exact} symmetry of the {\it
complete} $in$-state. Since the $in$-state
also describes the in-falling matter forming the black hole, this
is clearly not the case. We will return to this point later.

\newsection{Gravitational Back Reaction.}

This brings us to the question: how important are the gravitational
corrections? In particular, can the consequences be large enough to
avoid that information gets lost inside the black hole? Unfortunately,
there is not yet enough known about quantum gravity to definitely answer
this question. We can however, with some reasonable assumptions, try
to investigate what the possibilities are.

\newsubsection{Some general observations.}

To begin with,
let us recall another often mentioned argument for why,
in spite of the ultra-high energies involved, the
quantum gravitational effect should still be negligible in the
calculation of the Hawking state. Namely, one could argue
that in this  calculation it is unnecessary, or even wrong, to
trace the history of the out-going test particle all the way back
to ${\cal I}^-$, since intuitively the Hawking particles
emitted by the black hole begin their lives as virtual particles produced
by pair creation out of the vacuum near the horizon. Since this pair
production takes place after the black hole has already been
formed and settled into a ``steady state'', it
appears that the out-going radiation never had a chance to interact
with the in-falling matter. In other words,
the test-wave used in Hawking's derivation nor the virtual particles near
the horizon are physically detectable, and should according to this argument
have no measurable effects on the back ground geometry or anything else.

To test the validity of this argument, let us assume
that the black hole evaporation process satisfies some very basic
physical rules. Specifically, we will assume that the basic starting
points that were used in Hawking's
original derivation remain valid also after including the
gravitational back reaction. These physical principles can
be formulated as follows:

\medskip

($1$){\it There exists a quantum-mechanical time-evolution.}\\
In quantum mechanics we can describe the time evolution of a system using
either a Schr\"odinger or a Heisenberg picture. If we use the latter,
operators are time-dependent. The existence of a time-evolution implies
that all the operators at a given time $t_1$ must be expressible in terms
of the operators at any given earlier time $t_0<t_1$. We will not necessarily
require that this time evolution is strictly unitary: in principle we will
allow that certain operators are `destroyed' e.g. by the black hole
singularity, or transported into some other universe. However,
we do want to {\it exclude} the possibility that operators are created
out of nothing during the time evolution, since this would render the
quantum mechanical time evolution unpredictable in a clearly unacceptable
way. In other words, we will require that at any given time all observable
operators must have a past.

\medskip

($2$) {\it The initial asymptotic
data can be described in terms of a free field Fock space.}\\
This is a standard postulate of the LSZ asymptotic theory, and
assumes that in the remote past all particles are well-separated from each
other. In this limit all (gravitational) interactions between the
asymptotic particles can be ignored.
Of course, free field theory will no longer be adequate to describe the
black hole formation and evaporation process, but we will assume that
our first postulate is satisfied at all times. In particular, this means that
{\it in principle} all physical observables at later times are
expressible in terms of these initial free fields $\phi_{in}$.

\medskip

($3$) {\it Energy is conserved.}\\
Asymptotically, the space-time metric reduces at all times to the flat
Minkowski metric, and thus we can define a conserved energy operator $E$
that keeps track of the energy carried in or out of the system by the
asymptotic in and out-going particles.
By our assumption ($2$) there exists a unique vacuum state
%with energy $E=0$
and all other states have positive energy.

\medskip

It is important to note that none of the above three physical assumptions is
(obviously) contradicted by the semi-classical picture of the
evaporation process. We will see, however, that they will give very strong
restrictions on the possible scenarios. In particular, it is clear that
the first two requirements, while not as strong, will become essentially
equivalent to an $S$-matrix assumption if we would add CPT invariance
as an additional postulate.\footnote{For more detailed discussions of
CPT invariance in relation with black hole evaporation see \cite{cpt}.}
It should therefore be mentioned
that there are perhaps ways in which one can imagine weakening one or more
of these assumptions, but we are not aware of any accurately
formulated, acceptable alternatives.
%We will therefore adopt the above three rules as starting hypotheses,
%and let us now see what they lead to.

One of the immediate consequences of the first physical requirement (1)
is that any observable that can be used to
distinguish asymptotic states (\ie\ the asymptotic modes $b_\omega$)
must be expressible in terms of operators defined at earlier times.
Hence, although the Hawking radiation may perhaps be thought of
as arising from pair creation near the horizon, this does {\it not}
imply that the corresponding creation and annihilation
  modes $b_\omega$ and $b^\dagger_\omega$ are pair created out of nothing.
If this were true this pair creation process itself would already
violate a basic principle of quantum mechanics.  Thus the above
mentioned argument, that one should not propagate the particle waves
all the way back to ${\cal I}^-$, is clearly inaccurate, at least
when we interpret this propagation as the quantum mechanical time
evolution of a Heisenberg
operator. In quantum mechanics, to evolve states forward in time, we need
to know how to evolve observables backwards in time.

The first two physical assumptions combined imply
that the out-going modes ($b_\omega$, $b^\dagger_\omega$)
can be expressed in terms of
the in-coming modes ($a_\omega$, $a^\dagger_\omega$).
Given the form of the black hole geometry, this indeed seems
a reasonable conclusion. It is furthermore
the same assumption that underlies Hawking's derivation, which
suggests that this relation between the $b$- and the
$a$-modes will in a suitable approximation take the form of
a Bogoljubov transformation as in (\ref{Bogol}).

\newsubsection{Energy conservation}

In most semi-classical models of black hole formation and evaporation
the collapsing matter is assumed to be in some semi-classical (or coherent)
state. It then appears to be a sensible procedure
to replace the operators associated with it by classical $c$-number
quantities. However, the above
argument indicates that this may no longer be a good approximation when the
energy that is gained or lost by the test-particle is comparable to
the mass $M$ of the black hole. The main problem indeed with treating the
black hole geometry and the in-falling matter as $c$-numbers is that
it makes energy conservation totally obscure. All quantum processes
take place in a time dependent classical background, which
becomes an (inexhaustible) source of energy. This causes the
problem of the diverging frequencies.

In the following, we will therefore try to develop a formulation in
which we treat (part of) the in-falling matter quantum mechanically.
To do this exactly would of course be difficult, but as a reasonable
first approximation we will use Hawking's formulas as a starting point,
while replacing all quantities that depend on the background geometry
by {\it operator-valued} quantities. In other words, we will continue to
work with the relation (\ref{Bogol}) between the $in$- and $out$-modes,
but with the Bogoljubov coefficients replaced by suitable quantum
operators, that act on the Hilbert space of the in-falling matter.
Equation (\ref{Bogol}) will thus no longer be a linear relation but a
highly {\it non-linear operator
identification}, expressing the $out$-modes in terms of the $in$-modes.

It will become clear that in this framework one automatically includes
an important part of the interaction between the out-going
modes and the in-falling matter. Moreover, it will enable us to
formulate energy conservation as a meaningful and precise requirement.
Namely, we may imagine taking the commutator of the energy operator $E$ with
the operator valued Bogoljubov transformation  (\ref{Bogol}), and
require that we get the same answer on the left- and the right-hand-side.
The physical interpretation of this requirement is that all energy that
the test-wave gains in propagating back to ${\cal I}^-$, should come
from the matter forming the black hole.
We thus deduce that the coefficients
$\alpha_{\omega\xi}$ and $\beta_{\omega\xi}$ must carry energy $E=\xi-\omega$
and $E=\xi+\omega$, respectively.

{}From the classical expression (\ref{coeff})
of the coefficients we see that the energy balance can indeed
be restored while keeping essentially the same expression,
by giving parameter $\Tau_0$ a non-trivial commutation relation with
$E$ as follows
\be
[E, \Tau_0] = i.
\ee
so that
\be
\lbrack E,\alpha_{\omega\xi}\rbrack=i{d\over d{\Tau_0}} \alpha_{\omega\xi}
= (\xi-\omega) \alpha_{\omega\xi}
\ee
and similarly for $\beta_{\omega\xi}$. Although the precise meaning
of this observation can become fully clear only in a more complete quantum
theory of black holes, it certainly suggests that in understanding the
quantum back reaction of the Hawking radiation one should take into
account the operator character of $\Tau_0$. To illustrate this
point, we will now discuss one particularly important consequence
of this new insight.

%The significance of
%this fact will become even more clear in a moment.

\newcommand{\const}{c\,}%{{2e}}

\newsubsection{Gravitational shift interactions.}

%\noindent
A crucial assumption in Hawking's derivation is that the incoming particles
described by $\phi_{in}(\Tau)$ with $\Tau >\Tau_0$ and the outgoing particles
described $\phi_{out}(\Time)$ form independent sectors of the Hilbert space,
and that the corresponding field operators commute with each other.
The underlying classical intuition is that the fields $\phi_{in}(\Tau)$
with $\Tau>\Tau_0$ will propagate into the region behind the black hole
horizon, and thus become unobservable from the out-side. However,
this intuition
ignores the important fact that the in-falling particles in fact {\it do}
interact with the out-going radiation, because they slightly change the
black hole geometry. In the spherically symmetric theory, this change
in the geometry is represented by a small shift in the black hole mass $M$
of and the time $\Tau_0$ at which the black hole horizon was formed.

Consider a spherical shell of  matter with energy $\delta M$ that
falls in to the black hole at some  late time $\Tau_1$.
The Schwarzschild radius will then increase slightly with an amount
$2\delta M$, and the time $\Tau_0$ will also change very
slightly. A simple calculation shows that\footnote{Here, $c$ is a constant
of order one and happens to be
equal to $4e$}
\be
\label{deltatau}
\delta\Tau_0=-\const \delta M e^{-(\Tau_1-\Tau_0)/4M}
\ee
At first it seems reasonable to ignore this effect as long as the
change $\delta M$ is much smaller than $M$.
However, in view of our preceding discussion on energy conservation,
it may be a good idea to study this point somewhat closer.

The exponential $\Tau$-dependence that occurs in the formula
(\ref{deltatau}) is typical of black holes and has to do with the
diverging red-shift.
This time it helped in our favour because it exponentially suppressed
the effect on $\Tau_0$ of the in-going matter. But in other physical
quantities it is easy to get exponentially growing factors that enhance
physical effects that seemed to be unimportant at first.
For example, the variation in $\Tau_0$,
although very small,  has an enormous effect on the wave-function
$\phi_{out}(\Time)$ of an out-going particle.
By combining (\ref{repar}), (\ref{ttau}) and (\ref{deltatau})
one easily verifies that as a result of the in-falling shell,
the outgoing particle-wave
is delayed by an amount that grows rapidly as a function of $\Time$
\be
\label{deltapsi}
\phi_{out}(\Time)\ra \phi_{out}(\Time-
4M\log(1-{\const} {\delta M \over 4M} e^{(\Time-\Tau_1)/4M})).
\ee
Notice that even for a very small perturbation $\delta M$  the argument of
the field $\phi_{out}$  goes to infinity after a finite time
$\Time_{lim} -\Time_1 \sim -4M\log (\delta M/M)$.
The physical interpretation of this fact is
that a matter-particle that is on its way to reach the asymptotic
observer at some time $\Time>\Time_{lim}$ will,
as a result of the additional in-falling shell, cross the event-horizon and
be trapped inside the
black-hole horizon\footnote{This same calculation
is often used to show that "white holes" are unstable under small
perturbations.}.
This implies that the asymptotic wave-function of an {\it individual particle}
is {\it very sensitive} to the gravitational back-reaction. To see what this
means for the {\it collective} state of the outgoing radiation is clearly a
much more subtle matter. In fact, it can be shown that the transformation
(\ref{deltapsi}) is an approximate symmetry of the Hawking state, and this is
undoubtedly the reason why it is usually not taken in to account.
However, as we will show in the remainder of this section, the fact that
the gravitational back reaction is important for individual particles is
sufficient to substantially change the usual semi-classical picture.

\newsubsection{The exchange algebra between $in$ and
$out$-fields.}

How does one take the effect (\ref{deltapsi}) into account?
At this point we come back to our idea that the parameter
$\Tau_0$ is not just as a classical number but should be
treated as a quantum
operator.  To make this more concrete,  let us divide up the
in-falling matter
in a classical piece plus a small quantum part that is described
in terms of a quantum field $\phi_{in}(\Tau)$.  Of course,
$\Tau_0$ is mainly determined by the classical in-falling matter, but in
addition it has a small quantum piece. Using (\ref{deltatau}) we find
 \be
\Tau_0 =\Tau_0^{cl} - \const \int_{\Tau^{cl}_0}^\infty \!\! d\Tau \,
e^{(\Tau^{cl}_0-\Tau)/4M} T_{in}(\Tau)
\ee
where $T_{in}(\Tau)$ denotes the stress-energy tensor of the $\phi_{in}(\Tau)$
with support $\Tau>\Tau^{cl}_0$.
{}From here on the calculation is simple, straightforward and unavoidable.
The only additional ingredient we need is  the same
fundamental relation (\ref{repar}), that formed the starting point of
Hawking's derivation. The only difference is now that in the reparametrization
(\ref{ttau}) we include the seemingly negligible quantum contribution.
Our goal is to calculate  the algebra of the outgoing field
$\phi_{out}(\Time)$ for late times with the in-coming field
$\phi_{in}(\Tau)$ for $\Tau>\Tau^{cl}_0$. First we compute
\be
%\label{shift}
\lbrack \Tau_0,\phi_{in}(\Tau)\rbrack = -
i\const \exp((\Tau^{cl}_0-\Tau)/4M){\partial_\Tau}\phi_{in}(\Tau),
\ee
or equivalently
\be
\label{shift}
e^{i\xi\Tau_0}\phi_{in}(\Tau)e^{-i\xi\Tau_0}=\phi_{in}(\Tau-4M\log(1-{\const\over 4M} \xi e^{(\Tau^{cl}_0-\Tau)/4M})),
\ee
where we simply used the fact that the stress-tensor generates coordinate
transformations.

We can now compute the exchange algebra between the $in$ and
$out$-fields
by combining (\ref{repar}), (\ref{ttau})
 and (\ref{shift}).  One finds the following result
\be
\label{ex}
\phi_{out}(\Time)\phi_{in}(\Tau)=
\exp(i \const e^{(\Time-\Tau)/4M}\partial_\Time\partial_\Tau)
\phi_{in}(\Tau)\phi_{out}(\Time),
\ee
which is valid at for $\Tau>\Tau_0^{cl}$.
This exchange algebra is the quantum
implementation of the gravitational back-reaction (\ref{deltapsi}) of the
in-falling matter on the out-going radiation.
An equivalent
physical interpretation of this non-local algebra is that it represents
a gravitational shock-wave interaction between the incoming and out-going
matter waves \cite{thooft}.

We want to emphasize that to derive this result we did not need any
assumptions other than those already made in the usual derivation of
Hawking evaporation.
%\footnote{The assumptions (1), (2) and (3)
%were introduced in the previous subsection only for the sake of the
%argument, but are not essential in the derivation (\ref{ex}).}
The only extra ingredient that
we took into account is the small quantum contribution to $\Tau_0$. In this
sense the relation (\ref{ex}) appears to be unavoidable
and independent of which scenario one happens to believe in.

\newsubsection{Some physical consequences.}

Of course, much work is needed to analyze the precise
physical consequences of this algebra,
but at this point it is clear that
the presence of these large commutators
implies that the standard semi-classical
picture of the black hole evaporation process needs to
be drastically revised. In particular,
it tells us that, due
to the quantum uncertainty principle, we should be very careful in
making simultaneous statements about the in-falling and out-going fields.
Mathematically, the Hilbert space of the scalar-fields
on a Cauchy surface as drawn in figure 1 does {\it not}
decompose into a simple tensor
product of a Hilbert space inside the black hole and one out-side.
Instead, in view the exponentially non-local nature of the commutator
between the $in$ and $out$-fields, it is clear that
the $out$ Hilbert space is not even approximately independent of
the Hilbert space of the in-falling matter!

It should be pointed out, however,
that the exchange algebra (\ref{ex}) in itself
can not be responsible for the complete transfer of information to
the out-region, since it only depends on a single quantum number $\Tau_0$
of the in-going matter. In fact, as we will show explicitly in section
4.4, the above algebra is perfectly consistent with Hawkings
analysis, and can even be used to rederive his results in this new setting.
On the other hand, the algebra (\ref{ex}) has the important consequence
that it introduces large non-local quantum effects, that may bring
other (strong coupling) effects into the picture. In relation to
this it is
important to note that, in deriving the  algebra (\ref{ex}) for
$\phi_{in}(u)$ we excluded the $in$-region $\Tau \leq \Tau^{cl}_0$,
that is mapped in (\ref{repar}) on the asymptotic $out$-region.
In this region the algebra of $in$ and $out$-fields contains an
additional term, which describes the direct propagation of
the modes. In the leading semi-classical limit
\be
\label{refll}
\lbrack \phi_{out}(\Time),\phi_{in}(\Tau)\rbrack = i\theta(\Time-\Time(\Tau))
%+\ldots%\mbox{corrections}
\ee
where $\Time(\Tau)$ denotes the diffeomorphism given in (\ref{ttau})
that maps the out-interval onto the interval $\Tau<\Tau^{cl}_0$.
%where the dots represent the corrections that
%are due to gravitational back reaction.
This direct interaction between $in$ and $out$ modes is clearly
capable of transporting detailed information. The question that should be
studied further is whether the (yet to be found)
complete interaction between the $in$ and $out$
modes that combines the exchange relation (\ref{ex}) and the
direct propagation (\ref{refll}) is
sufficient to transport {\it all} the in-going quantum numbers
back out-wards into the Hawking radiation.

To illustrate why this is {\it in principle} possible, let us make
the following simple observation. We want to know whether the
out-going observables cover the complete collection of incoming
fields. To begin with,
it seems reasonable to assume that via the direct propagation
(\ref{refll}) the out-going operators contain at least all incoming
operators inside the region $\Tau<\Tau^{cl}_0$. This is nothing
new and in complete accordance with the usual semi-classical reasoning.
The important new consequence of the algebra (\ref{ex}), however,
is that it appears to hand as an additional peace of information to
the outside observer the quantum operator
$$
P_{in}= \int_{\Tau^{cl}_0}^\infty d\Tau e^{(\Tau^{cl}_0-\Tau)/4M} T_{in}(\Tau)
$$
that acts on the incoming fields at $\Tau>\Tau^{cl}_0$. At first sight
this may not seem such a big deal, since this is just one single quantum
number. It may therefore come perhaps as a surprise that, if both
these seemingly innocent assumptions are indeed correct,
then this immediately implies that the out-going modes
must carry {\it all} in-going information.
Namely, one should not think of $P_{in}$ as an operator
measuring just one quantum number (in fact, it has no eigen states),
but rather as an operator that {\it generates} a certain type of
translations on the interval $\Tau>\Tau_0^{cl}$.
It then becomes immediately evident that the collection of all
$in$-fields in the interval $\Tau<\Tau^{cl}_0$ {\it together} with the above
operator $P_{in}$ form a complete basis of operators for the
incoming Hilbert space!

Finally, we note that the algebra (\ref{ex}) has the surprising
property that it is {\it symmetric} between the $in$ and $out$-fields,
although the derivation certainly {\it looked} asymmetric. This
suggests that a complete description of the black hole
formation and evaporation process, that properly incorporates
this interaction, should also be symmetric between the $in$ and $out$
fields.

\newsection{Quantum Back Reaction in 2d Dilaton Gravity}

In this section we shall illustrate the above discussion
by focusing on the specific example of $d=2$ dilaton
gravity \cite{cghs}-\cite{hv}.
It is generally agreed that the physics of this
model is closely analogous to what happens in four dimensions
and, indeed, advocates on all sides of the black hole paradox
have skillfully used the model to strengthen their case. We
cheerfully joined this tradition in \cite{vv}-\cite{svv}, where we proposed
a specific $S$-matrix description of quantum dilaton gravity.
Here we will briefly review this approach, and show how it
naturally leads to the same type of non-local interactions
between the $in$ and $out$-fields as discussed above for the
$s$-wave reduction of the four-dimensional theory.

\newsubsection{Free field formalism}

Below we briefly review the analysis presented in \cite{vv}-\cite{svv}.
We will actually be a little bit more general by allowing
a general number $N$ of matter fields flavors and by
considering a more general class of boundary conditions
\cite{cv}.
Our intention in this section is to illustrate the points made
in the previous section in this specific situation. In particular,
it will become clear that the basic principles (1), (2) and (3)
proposed in section 3.1 are indeed realized in this simple model, and that
their implications can be understood.

The starting point of our analysis is the following
action for dilaton gravity coupled to $N$ massless scalar fields
in the conformal gauge \cite{cghs} \cite{cv}
\ba
S &=& \frac{1}{\pi} \int d^2 x \, \left[
(2 e^{-2\phi} (2\del_u\del_v \rho - 4\del_u\phi\del_v\phi
 + \lambda^2 e^{2\rho}) + \half \sum_{i=1}^{N} \del_u f_i \del_v f_i \right.
\nonumber\\[2mm] && \qquad \qquad \quad
\left. + {\textstyle{{N}\over{12}}}\, \phi \, \del_u\del_v (\rho-\phi)
  - {\textstyle{{N-24}\over{6}}} \del_u(\rho-\phi)\del_v(\rho-\phi) \right] \ .
\ea
This action contains a specific combination of 1-loop
correction terms, that was motivated in \cite{cv}.
Introducing
\be
\Omega(u,v) = e^{-2\phi} +
{\textstyle{{N}\over{24}}} \phi \ , \qquad
\rhohat = \rho - \phi \ ,
\ee
the equations of motion of this action can be written as
\ba
\del_u \del_v \, \rhohat &=& 0
\nonumber\\[2mm]
\del_u \del_v  \Omega + \lambda^2 e^{2\rhohat} &=& 0 \ .
\label{fe}
\ea
The stress-energy tensor takes the form
\be
T_{uu} = - 2 \del_u\Omega\del_u\rhohat + \del_u^2 \Omega
  - 2 \kappa ( \del_u \rhohat \del_u \rhohat - \del_u^2 \rhohat)
  + \half \sum_i \del_u f_i \del_u f_i \ ,
\ee
where we defined $\ka = \frac{N-24}{12}$.

The first of the field-equations (\ref{fe}) allows us to
introduce free field variables according to
\be
  e^{2\rhohat} = \del_u \X^+(u) \, \del_v \X^-(v) \ .
\ee
The general solution of the second equation in (\ref{fe})
is then given by
\be
  \Omega =
  -\lambda^2 \X^+(u) \X^-(v) + \omega^+(u) + \omega^-(v) \ .
\ee
We define variables $\PP_\pm$ according to
\be
\del_u \omega^+(u) = \PP_+ \del_u \X^+
   - \frac{\ka}{2} \frac{\del_u^2 \X^+}{\del_u \X^+} ,
\quad
\del_v \omega^-(v) = - \PP_- \del_v \X^-
   - \frac{\ka}{2} \frac{\del_v^2 \X^-}{\del_v \X^-} \ .
\ee
These definitions are such that at the quantum level
the following commutation relations are valid
\ba
[ \del \X^+(u_1), \del \PP_+(u_2)] &=&
              2 \pi i \, \delta^\prime(u_1-u_2) \ ,
\nonumber\\[2mm]
[ \del f_i(u_1), \del f_j(u_2)] &=&
              2 \pi i \, \delta_{ij} \, \delta^\prime(u_1-u_2) \ .
\ea
The total stress-energy tensor, which takes the form
\be
\label{stress}
T_{uu} =
    \del_u \PP_+ \del_u \X^+
    + \frac{\ka}{2} \del^2_u \log(\del_u \X^+)
    + \half \del_u f_i \del_u f_i
\ee
(normal ordering is implied), satisfies a Virasoro algebra of
central charge $c=2-12\ka+N=26$.

The semiclassical theory defined by these equations
has been analyzed recently in \cite{cv}.
Here we shall briefly
sketch a full quantum treatment of the theory, slightly
generalizing the work of \cite{vv}-\cite{svv}, where we focussed on
the special case $N=24$.

\newsubsection{Physical operators and gravitational dressing}

In the quantum theory, states that are annihilated by the total
stress tensor (\ref{stress}) are called physical states. Physical
states that are at the same time descendants with respect to the
total stress-tensor are called spurious physical states; it can be
shown that such states decouple from the theory. For the physical
spectrum we are thus interested in the space of physical states
modulo spurious physical states.

%(In BRST language this would,
%of course, correspond to the cohomology of the nilpotent BRST
%operator.)

The analysis of the physical state condition in quantum dilaton
gravity is closely analogous to a similar analysis in light-cone
gauge critical string theory. Following this lead, one discovers
that the following oscillators create physical states
(called DDF states)
\be
\label{Ddf}
\alpha_i(\omega) = \int\! du\,
        (\lambda \X^+(u))^{i \omega} \partial_uf_i(u) \ .
\ee
Furthermore, there are additional physical operators
given by
\ba
\alpha^-(\omega) \is \int \! du \,
\left[
:(\lambda \X^+)^{1+i\omega} \del_u \PP_+(u):
%+ \lambda^2 (1+i\omega) (\lambda \X^+)^{i\omega} \right.
\right. \nonumber\\[2mm]
&& \qquad \left. + \frac{\lambda\hat{\kappa}}{2} %(1+\kappa)
(1+i\omega)
(\lambda \X^+)^{i\omega} \del_u \log(\del_u \X^+)
\right] \ ,
\ea
with $\hat{\kappa}=\kappa+1$.
These operators are found to satisfy the following algebra
\ba
[ \alpha_i(\omega_1), \alpha_j(\omega_2) ] \is
\delta_{ij}\, \omega_1 \, \delta (\omega_1+\omega_2)
\nonumber\\[2mm]
[ \alpha^-(\omega_1), \alpha_i(\omega_2) ] \is
- \lambda \, \omega_2 \, \alpha_i(\omega_1+\omega_2) \ .
\ea
A most remarkable relation is the following
\be
[ \tilde{\alpha}^-(\omega_1), \tilde{\alpha}^-(\omega_2) ] =
\lambda \, (\omega_1-\omega_2) \tilde{\alpha}^- (\omega_1+\omega_2) \ ,
\ee
where $\tilde{\alpha}^-(\omega) = \alpha^-(\omega) - \frac{\lambda}{2}
:\!\alpha_i \alpha_i\! :(\omega)$. This relation shows that
the difference of the generator $\alpha^-(\omega)$ and the
light-cone Virasoro generator
$\frac{\lambda}{2} :\!\alpha_i \alpha_i\! :(\omega)$
generates a {\it centerless} Virasoro algebra.
This immediately suggests that states created by the
$\alpha^-(\omega)$ are equivalent (modulo spurious physical states)
to states that can be created by using the oscillators
$\alpha_i(\omega)$ only. A more careful analysis confirms that
the states created by using only the oscillators
$\alpha_i(\omega)$ indeed form a complete basis of all physical
states. (This statement is the equivalent of the so-called
no-ghost theorem in critical string theory.) In \cite{svv} we proved
this for $N=24$ ($\kappa=0$) by using a finite volume regularization.
Since this proof only used the algebraic relations satisfied
by the oscillators $\alpha_i(\omega)$ and $\alpha^-(\omega)$,
it can be generalized immediately to the general case
$N\neq 24$ ($\kappa\neq 0$).

We have thus obtained a very convenient basis for a
description of the space of physical states from the
point of view of an observer at ${\cal I}^-_R$, who
uses $\tau_+ = \lambda^{-1} \log(\lambda \X^+)$ as time
variable. A similar basis,
in terms of oscillators $\beta_i(\omega)$ appropriate
to an asymptotic $out$-observer at ${\cal I}^+_R$ can be
constructed by interchanging the role of the coordinates
$\X^+$ and $\X^-$. It is appropriate
to call the oscillators $\alpha_i$ `dressed oscillators',
since they correspond to `bare' oscillators of the
matter fields $f_i$ that are dressed with dilaton
gravity excitations (represented by $\X^+,\X^-$)
that describe the back-reaction of the matter excitation
on the dilaton gravity system. Our formalism
thus incorporates this back reaction in an explicit,
algebraic way that avoids the semi-classical notion of
a background dilaton gravity configuration. In the
next subsections we will show how this algebraic
treatment of the back-raction problem can be used
to derive Hawking radiation in the $out$-state and to
investigate the quantum gravitational corrections to
the semi-classical result.

\newsubsection{Boundary condition.}% in the Strong Coupling Region}

In the formalism for quantum dilaton gravity that we
described sofar, the left and right-moving degrees of freedom
are completely independent: incoming left-moving signals
propagate freely to ${\cal I}_L^+$ and outgoing right-movers
have their origin at ${\cal I}_L^-$. We are, however, interested
in a quantum mechanical theory that describes the evolution
of initial left-moving data on ${\cal I}_R^-$ to right-moving
signals on ${\cal I}_R^+$, in analogy with the 3+1-dimensional
theory. We shall now describe how such a
theory can be obtained by specifying a boundary condition in
the strong coupling regime.

Let us first use the classical theory to motivate the specific
boundary condition we will choose. Classically we can pick a
large but constant value of the dilaton field, and require
the line where this value is attained to be the boundary of the
two-dimensional world.\footnote{A boundary condition of this type
was first introduced by Russo et al in \cite{rst}.}
If we choose a physical coordinate system $(x^+,x^-)$
in which the rescaled metric $d\tilde{s}^2 = e^{-2\phi}ds^2$ (which
is flat everywhere) takes
the form $d\tilde{s}^2 = dx^+dx^-$, it
can be shown that the resulting equation of motion
for the boundary $(\xx^+,\xx^-)$ reads as follows
\be
\label{bound}
-{m\over 2}\sqrt{\partial_\mp \xx^\pm} \pm \lambda^2 \xx^\pm +
     p_\mp(\xx^\mp) = 0 \ ,
\ee
where
\be
\label{ppdef}
{p}_\pm(\xx^\pm) =
  \pm \int\limits_{{\xx^\pm}}^{\pm\infty} \!\! dx^\pm\, T_{\pm\pm}
\ee
denotes the integrals of the in- and out-going momentum flux.
Note that the $x^\pm$ coordinates are related to the standard
asymptotic coordinates $r$ and $t$
via $x^\pm = \pm \lambda^{-1} \exp(\lambda(r \pm t))$.

The total system of matter and boundary only describes a well-defined
dynamical system if one restricts to field configurations below a
certain critical energy flux. As an example, we consider
the classical boundary equation (\ref{bound})
when the incoming wave is a shock wave located at $x^+ = q^+$,
with amplitude $p_+$
\be
\label{shock}
T_{++}(x^+) = p_+ \delta(x^+-q^+)\ .
\ee
As long as the total energy  $E=p_+q^+$ carried by the pulse is smaller
than ${m^2 \over 4\lambda^2}$, the boundary trajectory is time-like
everywhere and given by
\be
\label{bounsh}
(\lambda^2 \xx^- - p_+)\xx^+ =
- \frac{m^2}{4\lambda^2}
\ee
for $\xx^+ < q^+$ and
\be
\label{bounsho} \xx^- (\lambda^2\xx^++p_-) =
- \frac{m^2}{4\lambda^2}
\ee
with
\be
\label{pmin}
p_-= \frac{\lambda^2 p_+ q^+}{{m^2 \over 4\lambda^2q^+} - p_+}
\ee
for $\xx^+ > q^+$.
The typical form of this boundary trajectory is depicted in fig 3a.
Note that the mirror point behaves as a particle with
negative rest mass: when the shock wave hits it, it does not bounce
back to the left but in the opposite direction to the right.

In case $p_+q^+ > {m^2 \over 4\lambda^2}$ then the solution to
the equation (\ref{bound}) cannot be time-like every where, but
turns space-like for $\xx^+ > q^+$ (see fig 3b).
So in this regime it is no longer classically consistent to treat
the boundary as a reflecting mirror. Obviously, this situation
precisely corresponds to the formation of a black hole.
This example suggest that the following inequality
\be
\label{ineq}
p_+(x^+) < {m^2\over 4\lambda^2 x^+}
\ee
for all $x^+$, with $p_+(x^+)$ as defined in (\ref{ppdef}),
is a necessary and possibly sufficient criterion for the
incoming energy flux to ensure that the classical boundary
remains timelike. Note that this inequality does not imply
any specific {\it local} upper bound on the energy flux
$T_{++}$.

\begin{center}
\leavevmode
\epsfysize=4.5cm
\epsfbox{figu3.eps}
\end{center}
\noindent
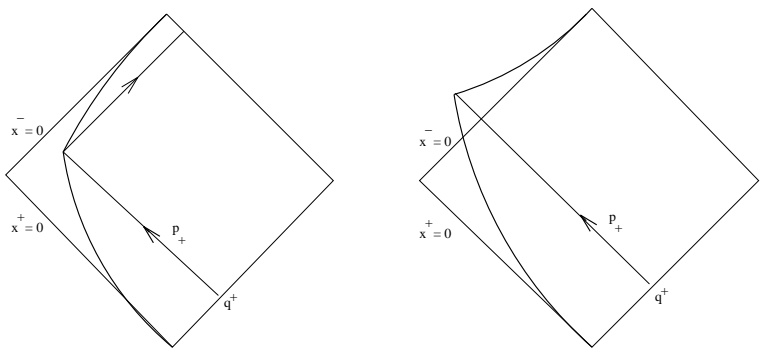
{\small Fig 3a and 3b. Schematic depiction of the classical boundary
trajectory for a sub-critical (left) and a super-critical (right)
shock wave.}

Let us now describe how the above boundary condition translates
to the quantum theory described in the previous subsection.
We choose the $(u,v)$-coordinate system in such a way that
the boundary becomes identified with the line $u=v$, and denote
the parameter along this boundary by $s$.
We require that the dilaton field takes a large and constant
value along the boundary, such that
\be
\label{com}
\partial_s \Omega = 0 \ .
\ee
In terms of the $\X$ and $\PP$-fields this condition reads
\be \label{bon}
\partial_s \X^+ (\lambda^2 \X^- - \PP_+)
  + \partial_s \X^- (\lambda^2 \X^+ + \PP^-)
  + {\kappa\over 2} \partial_s \log(\partial_s \X^+ \partial_s \X^-)
  = 0.
\ee
The above boundary condition is coordinate invariant and this
allows us to impose the additional constraint that the
gravitational and matter components of the energy momentum
flux each separately reflect off the boundary. So
(\ref{bon}) is supplemented with the condition
\be
\label{hct}
T^{g}_{uu} =T^{g}_{vv}
\ee
where (compare with (\ref{stress}))
\ba
T^g_{uu} &=&
    \del_u \PP_+ \del_u \X^+
    + \frac{\ka}{2} \del^2_u \log(\del_u \X^+)
\nonu
T^g_{vv} &=&
    - \del_v \PP_- \del_v \X^-
    + \frac{\ka}{2} \del^2_v \log(\del_v \X^-) \ .
\ea
The two equations (\ref{bon}) and (\ref{hct}) combined specify
the precise reflection condition that relates the incoming
canonical variables $(\X^+,\PP_+)$ to the outgoing canonical
variables $(\X^-,\PP_-)$.

We would like to make manifest
that this relation defines a canonical transformation.
To this end, we should write a generating functional $S[\X^+,\X^-]$
of the coordinate fields, such that the momenta $P_\pm$ defined by
\be \label{gena}
\PP_{\pm} = {\delta S[\X] \over \delta \partial_s \X^{\pm}}
\ee
identically solve the boundary equations (\ref{bon}) and
(\ref{hct}). This condition results in a set of functional equations for
$S[\X]$ that can be solved explicitly. The form of the solution is
unique, once we fix the constant value of $\Omega$ along the boundary.
If we set $\Omega(\X^+,\X^-) = {m^2 \over 4\lambda^2}$,
then the generating functional $S[\X]$ takes the following form
\ba
S[\X] &=& m \int ds
\sqrt{\partial_s {\X}^+ \partial_s{\X}^-}
- \lambda^2\int\! ds \, \X^+\partial_s {\X}^-
\nonumber\\[2mm]
&& + {\kappa\over 2}
\int ds \log(\partial_s \X^+) \partial_s \log(\partial_s \X^-) \ .
\label{genfun}
\ea
Using (\ref{gena}) we then obtain the following relations
between the in-going variables $(\X^+,\PP_+)$ and the
outgoing variables $(\X^-,\PP_-)$\footnote{This equation
in fact requires a suitable normal ordering prescription. In the
semi-classical limit such corrections will not be important.}
\be
\label{newmom}
\pm {m\over 2}  \sqrt{\partial_s \X^+ \partial_s \X^-} +
\partial_s \X^\pm(\lambda^2 \X^\mp \mp \PP_\pm) + {\kappa \over 2}
\partial_s \log(\partial_s \X^\mp) = 0 \ .
\ee
As was shown in \cite{cv}, the relations (\ref{newmom})
can be used to derive the following semi-classical equation
of motion for the coordinates $(\xx^+,\xx^-)$ of the boundary curve
\be
\label{eoma}
-{m\over 2}\sqrt{\partial_\mp \xx^\pm} \pm \lambda^2 \xx^\pm +
p_\mp(\xx^\mp)
\pm{\kappa\over 2}\partial_\mp \log \partial_\mp \xx^\pm = 0 \ ,
\ee
where the physical fields $p_\pm(\xx^\pm)$ are identified with
the integrals (\ref{ppdef})
of the respective components of the matter stress tensor.

The equation (\ref{eoma}) can be compared to the equation (\ref{bound})
which can be derived in classical dilaton gravity. This
comparison shows that in general the classical boundary trajectory
receives quantum corrections.
At the quantum level, the dynamics of the boundary
curve is affected by the fact that the stress-energy
flux $T_{\pm\pm}$ does not vanish in the physical
vacuum defined by an asymptotic observer. In particular,
this implies that, in order to have a {\it time-like}
vacuum boundary curve, the parameter $m$ has to
satisfy the inequality
\be
m^2 > 8\ka \lambda^2 \ .
\ee
We refer to \cite{cv} for further discussion of this point.

Before we continue, we would like to reflect on the profound
consequences of the equations (\ref{newmom}). After the
boundary condition has been implemented, the operators
$(\X^+,\PP_+)$ and $(\X^-,\PP_-)$ all act in the same Hilbert space
and have non-trivial commutation relations. In particular,
the out-going coordinates $\X^-$ no longer commute with the
in-going coordinates $\X^+$. In the special case $\kappa=0$,
one can explicitly compute this commutator, which takes the form
(again modulo normal ordering effects)
\be
\label{comu}
[ \X^\pm (u_1) , \X^\mp (u_2) ] =
   {2\pi i \over \lambda^{2}} \, e^{{-\lambda^2 \over m} \int_{u_1}^{u_2}\!
du \sqrt{\partial_u {X}^+ \partial_u{X}^-}}\ .
\ee
The function on the right-hand side is equal to $2\pi i \lambda^{-2}$ for
$u_2>u_1$ and dies off exponentially with the physical distance
for $u_1>u_2$. In the limit $m=0$, which was considered in detail
in \cite{svv}, the above algebra reduces to the standard
free field commutation relation
\be
\label{commu}
[ \X^\pm (u_1) , \X^\mp (u_2) ] =
    {2\pi i \over \lambda^{2}} \, \theta(-u_{12}) \ .
\ee
In the next subsection
  we will show that this non-trivial
commutator between the physical coordinates $\X^+$ and $\X^-$
leads to an exchange algebra between physical $in$- and $out$-fields
very similar to the one discussed in section 3. We will further
demonstrate that, among other
things, it directly implies the existence of Hawking radiation.

\newsubsection{The exchange algebra and Hawking radiation.}

In this subsection we focus on the special case of $N=24$ matter fields,
so that $\kappa=0$. Let us introduce the following physical operators
\be
\label{Ahat}
\widehat{A}_i(p_+)=\int\! du \,  e^{-ip_+X^+(u)}\partial_uf_i(u)%\nonu
%\widehat{B}_j(p_-)\is\int\! du\, e^{-ip_-X^-(u)} \partial_uf_j(u),
\ee
that create in-coming  %and out-going
particles
with a definite Kruskal-momentum
$p_+$. %and $p_-$ respectively.
These operators can be expressed
as linear combinations of the energy eigenmodes
$\alpha_i(\omega)$. % resp. $\beta_j(\omega)$.
For example, for $p_+>0$ we have
\be
\label{Alpha}
\widehat{A}_i(p_+) = -i \int\! d\omega\, e^{-{\pi\over 2}\omega}
 \Gamma(-i\omega)
 \Bigl({p_+\over\lambda}\Bigr)^{i\omega} \alpha_i(\omega).
\ee
This equation, which can also be read as the definition of
$\widehat{A}_i(p_+)$, shows that these Kruskal modes are in fact somewhat
singular operators, because
they contain $\alpha_i(\omega)$ modes of arbitrarily
high frequency. In the following we will mostly ignore this singularity,
as it will not affect the main conclusions.

Using the operators  $\widehat{A}_i(p^+)$ we can build a more or
less realistic
incoming state in which the  matter is localized in a finite time interval
$x^+_0<x^+<x^+_0+\Delta x^+$ and carries a large total
energy $E \pm \Delta E$. This $in$-state may be represented
as a sum of eigenstates of the total Kruskal-momentum with
eigenvalues  concentrated around $\PP_+=E/x^+_0$,
and each of these eigenstate  can be constructed
by acting with a string of  $\widehat{A}_i(p^+)$-operators on the vacuum.
Thus a typical incoming state is a linear combination of states of the form
\be
\label{psi}
|\psi\rangle =
\prod_n  \widehat{A}_{i_n}(p_n^+)\vac.
\ee
In a Heisenberg picture the $out$-state is given by this same expression,
but to interpret this state physically we have to know how to write it in
terms of the $out$-fields.

The idea is now to analyse this state by acting with the out-going modes and
to try to commute this modes through the $\hat{A}$-operators.
For the outgoing modes we choose to work in a coordinate representation
\be
\label{fout}
f^{(out)}_i(x^-) = \int {d\omega\over \omega}
(\lambda x^-)^{i\omega} \beta_i(\omega)
\ee
where $x^-$ is a $c$-number and  $\beta_i(\omega)$ the operator
(compare with (\ref{Ddf})
\be
\label{Bo}
\beta_i(\omega) = \int dv (-\lambda \X^-(v))^{-i \omega} \partial_vf_i(v) .
\ee
{}From the commutation relation  (\ref{comu})
between the $\X^+$ and $\X^-$-fields
%and discuss its consequences for the algebra of
%physical $in$-and $out$-fields.
%First we note that (\ref{commu}) can be integrated to
%\be\label{comm}[\X^-(u_1),\X^+(u_2)] = 2\pii\lambda^{-2} \theta(-u_{12}).\ee
%We will now show that this physical effect is the origin of
%the Hawking radiation within the context of dilaton gravity.
it is straightforward to derive the exchange properties
of the $in$ and $out$ fields (see \cite{svv}). The details of this
algebra depend on the particular boundary condition that is chosen
in the strong coupling region, but there is one particular universal
feature that is independent of this choice. Namely, for
any choice of the parameter $m$, the two coordinate
fields $\X^+$ and $\X^-$ have the following commutator in the
region  $u_1<u_1$
\be
\label{comus}
[\X^+,\X^-]={2\pi i\over \lambda^2}.
\ee
This shows that the $in$-fields $f^{(in)}$, since they depend
$\X^+$, are capable of shifting the argument $\X^-$ of the
$out$-fields $f^{(out)}$.  This
is exactly the same gravitational effect we discussed before
in section 3, and corresponds physically to the fact that the
$in$-coming fields interact gravitationally with the $out$-fields
even before they disappear into the black hole and/or reflect off the
boundary.  We would now like to study the physical effect of this
interaction.

Since this interaction essentially takes place in the region $u_1<u_2$,
we will now simplify the discussion and for the moment work
with the simplified commutation relation (\ref{comus}).
We thus find the following exchange algebra of
$\widehat{A}_i(p_+)$  and $f_j^{{(out)}}(x^-)$:
\be
\label{shft}
\qquad f_j^{{(out)}}(x^-) \, \widehat{A}_i(p_+) =
 \widehat{A}_i(p_+) \, f_j^{{(out)}}(x^-\!\! -\!{p_+\over \lambda^2}) ,
%\qquad \qquad x^- < x^-_0 .
%+ \tilde{R}_{ij}(p_+,x^-),
\ee
We explicitly see that the $in$-mode shifts the $out$-fields by an
amount proportional to the incoming Kruskal-momentum
$p_+$.  %Thus (\ref{shft}) incorporates the shift-interaction due to
%the shockwave of an incoming particle.
Note that this interaction
is essentially the same as that derived in the $s$-wave reduced Einstein
theory with $1/4M$ replaced by $\lambda^{-2}$.

If the coordinate $x^-$ were a normal Minkowski coordinate, ranging
from $-\infty$ to $+\infty$, a constant shift in $x^-$ would
have had no physical effect whatsoever: it could simply be absorbed
by shifting the Minkowski vacuum. In our case, however, it is
crucial that $x^\pm$ parametrizes only a Rindler wedge $\pm x^\pm>0$
and that the vacuum of the $f$-fields is defined accordingly in
terms of the Rindler type modes $\alpha(\omega)$ and $\beta(\omega)$.
We will now show that the
resulting distortion of the out-modes is responsible for the
production of Hawking radiation.

To determine how the out-going vacuum state is affected by the coordinate
shift (\ref{shft}), let us rewrite the exchange algebra in terms of the
$\beta$-modes.  One finds
\be
\label{exch}
\beta_j(\omega) \widehat{A}_i(p_+) =  \widehat{A}_i(p_+)\int\!d\xi
\,{\rm B}_{\omega\xi}(p_+)\beta_j(-\xi)
\ee\
with
\be
\label{bcoef}
B_{\omega\xi}(p_+) =\Bigl({p_+\over\lambda}\Bigr)^{-i(\xi+\omega)}
{\Gamma(1-i\omega)\Gamma(i(\xi+\omega))\over\Gamma(1+i\xi)}.
\ee
The linear combination of $\beta$-modes
on the right-hand-side contains both creation- and
annihilation- operators. Consistency of the algebra (\ref{exch}) further
requires that these combinations again satisfy canonical commutation
relations, so we see that exchanging a $\beta_j(\omega)$-oscillator with
$\widehat{A}_i(p_+)$ leads to a Bogoljubov-transformation. Note that
the transformed modes occurring on the
right-hand side of (\ref{exch}) do not form a complete basis of all
$\beta$-modes, since they cover only the interval
$x^-<-\lambda^{-2} p_+$.  The exchange property (\ref{exch}) can
therefore in general only be used in one direction.

Finally, we can act on the state $|\psi\rangle$ given in (\ref{psi})
with the $\beta_j$-modes and repeatedly use (\ref{exch}) until we can act
on the vacuum. These manipulations are of course the direct quantum
counterpart of the standard semi-classical calculation.
The repeated use of the exchange-algebra describes the propagation of the
out-going particles through the
in-falling matter, while taking into account
the gravitational interaction between the two. Now  it is clear  from
(\ref{shft}) that in this procedure only the total
momentum $\PP_+$ plays a role, so the exchange relation between
the $\beta$-modes and the product $\prod_a\widehat{A}_{i_a}(p_a)$
is again of the same form (\ref{exch}). In this way we find that, just as in
the semi-classical calculation, the asymptotic out-state is
given by the Bogoljubov transform
(\ref{exch})-(\ref{bcoef}) of the vacuum.  At this point we can simply refer
to the standard
analysis \cite{hawking} to conclude that, in this approximation,
the $out$-state describes a constant out-going flux of
thermal Hawking radiation.

It is important to note that above we have of course
dealt with an idealized situation. In the first place we assumed that
$|\psi\rangle$ is an exact eigen state of $\PP_+$, and as
noted before, such states in fact do not exist.
If instead we would consider a state for which the in-coming energy is
bounded, we should find that the resulting black hole radiate for only a
finite amount of time. This can in fact be very directly seen in the
above algebraic derivation. The exchange algebra we derived for dilaton
gravity is namely
{\it energy preserving} in the precise sense that the total energy
carried by the $in$-modes plus that carried by the $out$-modes is the same
on both sides of the equation. This tells us that
every time a out-going particle is emitted, the in-falling matter
loses a proportional amount of energy, while it also gets shifted.
This process can continue for a long time, until most of the in-coming
energy has been radiated away.
We believe that by this time new (strong coupling)
effects will start to take place, while also the direct reflection
off the $r=0$ boundary will again become important. These effects
will produce corrections to the Hawking spectrum,
that in principle should be capable of transporting all information
to the out-side world.  The detailed physics of these processes
depends on the boundary conditions provided by the strong coupling
physics. A specific proposal for these boundary conditions for dilaton
gravity %(as described in section 4.3 with $\kappa=0$ and $m=0$)
has been worked out in detail in \cite{svv}.

\newsection{Concluding Remarks.}

To summarize, we have shown in this note that
gravitational interactions lead to non-trivial commutators
between the observables that measure the Hawking radiation and
the matter falling into the black hole.
To derive this commutator algebra (\ref{ex}) we did not need to
make any new assumptions other than those already made in the
Hawking's original derivation of black hole evaporation effect
\cite{hawking}.
A very important consequence of this result is that it calls into
question the standard
assertion that the Hilbert space of the scalar-fields
on a Cauchy surface as drawn in figure 1 decomposes into a simple tensor
product of a Hilbert space inside the black hole and one out-side.
Instead, our result supports the physical picture that there is
a certain {\it complementarity} between the physical realities
as seen by an asymptotic observer and by an in-falling
observer. Indeed, for the latter, the in-falling
matter will simply propagate freely without any perturbation, but he
(or she) will not see the out-going radiation. For the asymptotic
observer, on the other hand, the Hawking radiation is physically real,
%due to the non-local interaction between $in$ and $out$-fields,
while the in-falling matter will appear to evaporate completely before it
falls into the hole. With our result, the apparent discrepancy between
these observations can be explained by the fact that the two observers
use different {\it non-commuting} sets of observables to assign physical
meaning to the same quantum state. It is clear that this insight
will have important repercussions for the information paradox (see also
section 3.5).

As a further comment, we note that our
derivation of the commutator between the in-falling and out-going modes
can of course be generalized to other observers, by properly taking into
account the precise {\it operator} relation between the physical
coordinates
used by these observers. In this way one will only find strong coupling
effects when the relation between these coordinate systems becomes
very singular, such as in the case worked out in section 3.
Our reasoning does therefore not single out the horizon as a singular
part of space-time. It are only the {\it observables} that the
asymptotic observer uses to distinguish different physical
situations that become singular, since his (or her) coordinate system
degenerates there.  The
coordinate system of the in-falling observer, on the other hand,
is regular, so she (or he) will see nothing special happening
near the horizon.

Finally, we expect that our result may also be of importance
in relation with the entropy of black holes.
In \cite{thent} it was noted that a naive free field calculation of the
one-loop correction to the black hole entropy gives an infinite
answer. This infinity arises due to the diverging contribution
of states that are packed arbitrarily close to the horizon.
Our result suggests a possible remedy of this problem, because
it shows that in the coordinate system appropriate for this calculation
the in- and out-going fields no longer commute when they come
very close to the horizon. It is tempting to speculate that this
will effectively reduce the number of allowed states, and thereby
eliminate the diverging contribution in the entropy
calculation. We leave this problem for future study.

\medskip

\noindent
{\bf Acknowledgements.}

\noindent
We would like to thank U. Danielson and C. Stephens for stimulating
discussions. The research of K.S. is supported by DOE-Grant DE-AC02-76ER-03072,
and that of H.V. is supported by NSF Grant PHY90-21984 and the David
and Lucille Packard Foundation.
The research of E.V. is partly supported by an Alfred P. Sloan Fellowship,
and by a Fellowship of the Royal Dutch Academy of Sciences.

\end{document}